\documentclass[aps,prb,twocolumn,superscriptaddress, preprintnumbers]{revtex4-2}
\usepackage{latexsym}
\usepackage{amssymb}
\usepackage{graphicx}
\usepackage{amsmath}
\usepackage{bm}
\usepackage[colorlinks,
          linkcolor=black,
            citecolor=black,
            urlcolor=blue
           ]{hyperref}
\usepackage{verbatim}
\usepackage{mathrsfs}
\usepackage{extarrows}
\usepackage{comment}
\usepackage{mathtools,slashed}
\usepackage{soul}
\usepackage[toc,page]{appendix}
\usepackage[vcentermath]{youngtab}
\usepackage{multirow}
\usepackage[width=.75\textwidth]{caption}
\usepackage{atbegshi,picture}
\usepackage{lipsum}
\usepackage[normalem]{ulem}
\usepackage{soul}
\usepackage{xcolor}
\setstcolor{blue}   % 删除线颜色

\makeatletter
\renewcommand\@makecaption[2]{
  \par
  \vskip\abovecaptionskip
  \begingroup
   \small\rmfamily
    \begingroup
     \samepage
     \flushing
     \let\footnote\@footnotemark@gobble
     \@make@capt@title{#1}{#2}\par
    \endgroup
  \endgroup
  \vskip\belowcaptionskip
}
\makeatother

\makeatletter
\begin{document}

\title{Thouless pumps and universal geometry-induced drift velocity in multi-sliding quasi-periodic lattices}

\author{Zixun Xu}
\email{zixun.xu@sjtu.edu.cn}
\thanks{Corresponding author.}
\affiliation{Institute of Condensed Matter Physics, School of Physics and Astronomy, Shanghai Jiao Tong University, Shanghai 200240, China}

\author{Yuan Yao}
\affiliation{Institute of Condensed Matter Physics, School of Physics and Astronomy, Shanghai Jiao Tong University, Shanghai 200240, China}

\date{\today}

\begin{abstract}
Quantized Thouless pumps in periodic systems, set by Chern numbers or Wannier-center winding, is by now fairly well established, whereas its quasi-periodic extensions still require further clarification. 
Here, we develop a general quantitative paradigm for bulk Thouless pumps in continuous models with spacetime quasi-periodicity, applicable to arbitrary spatial dimensions. 
Within this framework, the bulk pumping turns out to be governed by an emergent long wave-length effective potential.
Based on this mechanism,
we obtain our main result a universal relation between topological drift and the geometry of quasi Brillouin zone. 
Reduced to periodic systems,
our result gives an explicit and compact formula which enables us to directly calculate Chern numbers by microscopic data.
These proposals are corroborated by simulations of one- and two-dimensional continuous moiré-type spacetime quasi-periodic lattices, which exhibit stable, localized, directional drift in excellent agreement with the theory.

\end{abstract}

\maketitle

\section{Introduction} Classifying and recognizing various quantum phases is a central issue in statistical and condensed matter physics.
Characterizations of gapped quantum phases by various quantized responses,
so-called topological invariants,
have attracted great efforts.
Thouless pumps~\cite{PhysRevB.27.6083},
as the dynamical analogue of the typical quantum Hall effect \cite{PhysRevLett.49.405}, are canonical examples of quantized transport: the charge transferred across the system during each adiabatic cycle corresponds to a topological invariant—the Chern number that is robust against local perturbations, inspiring broad interest as a proper realization of novel current standards~\cite{PhysRevLett.64.1812,RevModPhys.85.1421}. In the past decade, Thouless pumps have been studied in interacting~\cite{jurgensen2021quantized,nonlinearye,4d5s-n4gn,96f5-qszj}, non-Abelian~\cite{PhysRevLett.128.244302,sun2022non,sun2024two,PhysRevA.103.063518}, disorder~\cite{PhysRevA.103.043310,PhysRevA.101.052323,PhysRevLett.124.086602,liu2025interplay} and higher-symmetry protected systems~\cite{PhysRevB.100.245134,PhysRevResearch.2.022049,PhysRevLett.128.246602,PhysRevResearch.2.012009}. Quasi-periodic systems, in particular, have emerged as a key platform exhibiting exotic phenomena and rich topological transport behavior~\cite{PhysRevLett.109.106402,zilberberg2018photonic,lohse2018exploring,PhysRevLett.125.224301,PhysRevB.91.064201,nakajima2021competition,citro2023thouless}.
%Realizing perfectly quantized pumping in electronic materials is, however, challenging due to the need for exact band filling and strict adiabaticity. Experimental progress has thus focused on highly controllable platforms such as photonic lattices 
%\cite{kraus2012topological,jurgensen2021quantized,zilberberg2018photonic,wang2022two,sun2022non}, ultracold atoms~\cite{lohse2016thouless,nakajima2016topological,lohse2018exploring,nakajima2021competition} and acoustic metamaterials ~\cite{cheng2020experimental}.

In Thouless’s original model \cite{PhysRevB.27.6083}, the potential comprises two sliding commensurate sublattices. Extending this concept to multiple sublattices incommensurate in space and time leads to quasi-periodic structures. In periodic systems,
pumping can be described by the quantized displacement of the Wannier center or integer charge transfer over one adiabatic cycle. 
On the other hand, the study of spacetime quasi-periodic structures is limited due to
the infinitely many bands and unbounded periods,
which invalidate a conventional Brillouin zone and a well-defined Wannier center displacement. The pumping current density and the bulk-state drift velocity, as commensurability-independent characterizations, are therefore needed. Early studies of charge pumping driven by interlayer sliding in moiré superlattices related the pumped charge to sliding Chern numbers and recognized that the pump process is connected to the sliding of the moiré pattern \cite{PhysRevB.101.041410,PhysRevB.101.041112,PhysRevB.101.041113}. Building on this insight, a topological gap labeling based on the quasi Brillouin zones (qBZ) was developed for two-dimensional (2d) \cite{PhysRevResearch.4.013028,yoshii2023gap,jat2024higher} and three-dimensional (3D) \cite{PhysRevB.105.115410} systems. For the special case of a single sliding sublattice, the pumping current density is directly determined by the geometry of the qBZ. \cite{PhysRevResearch.4.013028,PhysRevB.105.115410}

Moreover,
multiple incommensurate modulations induce complicated interband couplings and nontrivial Landau Zener (Zener) tunneling ~\cite{wittig2005landau}. The study on bulk-state dynamics and  how it encodes the underlying topological invariants is still lacking, e.g.,
one-dimensional (1D) topological drift velocity was numerically found to be related to,
with a high precision,
averaged Chern number of populated bands~\cite{yang2024observation},
while the underlying mechanism of this empirical rule is unknown. A related observation in twisted bilayer graphene showed that flat-band electrons localize at the AA-stacking regions and are transported synchronously with the sliding moiré pattern \cite{PhysRevB.103.155410}, suggesting that bulk-state dynamics may be governed by a moving effective pattern or potential. Despite these advances, a systematic framework for spacetime quasi-periodic system beyond purely spatial (or temporal) cases and its extension to higher dimensions remain an open question.

In this paper, we introduce a general framework for bulk Thouless pumps in continuous models with spacetime quasi-periodicity. We extend the gap labeling formalism from a single sliding sublattice to multi-sliding cases in arbitrary spatial dimensions and obtain a general expression for the pumping current density. We show that the pumping process,
whether periodic or not,
is governed by an emergent long-wavelength effective potential.
Based on this mechanism,
we establish a universal relation between the topological drift velocity and geometry of qBZ.
Within the commensurate regime,
our results yield compact Chern number formula.
Furthermore,
we show that the previously observed averaged Chern-number phenomenon~\cite{yang2024observation} can be quantitatively proven by the effective-potential mechanism.
To confirm our proposal,
we perform numerical simulations of 1D and 2D continuous moiré-type spacetime quasi-periodic lattices. 
The results reveal stable, localized, and directional drift of an initially localized state,
in excellent agreement with the predicted drift velocity.

\begin{figure}[b]
\centering
\includegraphics[width=0.45\textwidth]{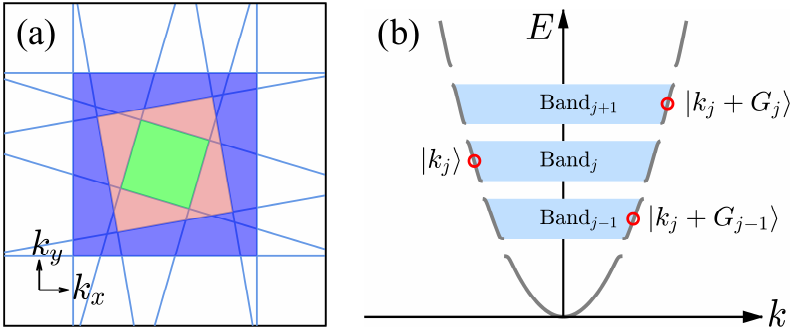}% Here is how to import EPS art
\caption{\label{fig.1}(a) Bragg planes of a 2D quasi-periodic lattice are shown as blue lines, and three qBZs are filled with different colors. 
(b) Band structure near $\mathrm{Band}_j$ in the nearly-free-electron approximation, with red circles marking the positions of three unperturbed plane wave states. 
}
\end{figure}

\section{Thouless pumps in d-dimensional lattices}
We consider the system governed by a dimensionless linear Gross–Pitaevskii (or
Schrödinger) equation,
\begin{equation}
    i \frac{\partial}{\partial t} \Psi = -\frac{1}{2}  \bm{\nabla} ^2 \Psi+V(\bm{r},t) \Psi.\label{eq:sde}
\end{equation}
The natural units $\hbar=m=1$ are adopted. To analyze the band structure of the system, we define the instantaneous eigenstates $\psi_{n}(r,t)$ and eigenenergies $E_{n}(t)$: $H(r,t)\psi_{n}(r,t)=E_{n}(t)\psi_{n}(r,t)$. Suppose the potential can be split into several sliding sublattices:
\begin{equation}
V(\bm{r},t)=\sum V_m \exp[i (\bm{b_m}\cdot \bm{r}-\phi_m t)]+\text{h.c.}.
\label{V(r,t)}
\end{equation}
Such a system is driven by multiple frequencies, with its commensurability determined by the set $\{\bm{b}_m, \phi_m\}$. In the commensurate case, e.g., $V(x,t)=\cos(x-t)+\cos(2x-3t)$, the system possesses well-defined spatial and temporal periods. Consequently, notions such as Brillouin zones and energy bands are well-defined, revealing quantized charge transfer and integer Wannier-centre displacement. In the incommensurate case, e.g., $V(x,t)=\cos(x-t)+\cos(\sqrt{5}x-\pi t)$, the system has no period in either space or time and exhibits quasi-periodicity. Thus the above notions are ill-defined, and we should use pumping current density and the bulk-state
drift velocity to characterize the pumping behavior instead. We present below a universal framework built on the qBZ geometry, regardless of commensurability.

In the nearly-free-electron (NFE) approximation, integer linear combinations of the reciprocal vectors determine a set of Bragg vectors, $\mathbf{G} = \sum_m n_m \mathbf{b}_m$, where $n_m\in\mathbb{Z}$ indicate the order of perturbation \cite{niu1986quantum}. 
Each Bragg vector $\mathbf{G}$ corresponds to a Bragg plane, which is the perpendicular bisector of $\mathbf{G}$ in momentum space. 
A Bragg gap opens at the Bragg plane under a generic perturbation. These Bragg planes form a network that provides a geometric partition of the momentum space. 
Each qBZ is defined as the region associated with a specific isolated group of energy bands, whose boundaries coincide with the Bragg planes corresponding to the upper and lower energy gaps of that band group \cite{PhysRevResearch.4.013028,PhysRevB.105.115410,yoshii2023gap,jat2024higher}, as illustrated in Fig.~\ref{fig.1}(a). If the system is  incommensurate in space, energy bands in the strict sense are ill-defined, yet we still refer to each such group as a band group. One can express the volume of a qBZ as
\begin{eqnarray}
S_{qBZ} &=& \sum_{{\alpha}=\{ \alpha_1,\alpha_2,\cdots ,\alpha_d \}}{\nu _{\alpha_1,\alpha_2,\cdots ,\alpha_d}}S_{\alpha_1,\alpha_2,\cdots ,\alpha_d},
\label{S_BZ}
\end{eqnarray}
where $S_{\alpha_1,\alpha_2,\cdots ,\alpha_d}\equiv\star(\bm{b}_{\alpha_1}\land \bm{b}_{\alpha_2}\land \cdots \land \bm{b}_{\alpha_d})$,
with $\star$ the Hodge star and $\land$ the exterior product. 
Each $S_{\alpha}$ corresponds to a fundamental Brillouin zone spanned by the selected set of $\{\bm{b}_{\alpha_1},\bm{b}_{\alpha_2},\cdots,\bm{b}_{\alpha_d}\}$. 
Here ${\alpha}$ denotes a $d$-element subset such that $\{\bm{b}_{\alpha_1},\bm{b}_{\alpha_2},\cdots,\bm{b}_{\alpha_d}\}\subset\{\bm{b}_m\}$, where the set $\{\bm{b}_m\}$ contain all wave vectors in Eq.~\eqref{V(r,t)}, and the summation $\sum_{\alpha}$ runs over all such subsets. 
For brevity, we write $S_{qBZ}=\sum_{{\alpha}}{\nu _{{\alpha}}}S_{{\alpha}}$. 
If the number of vectors in $\{\bm{b}_m\}$ is $K$, there are $C_K^d$ possible subsets $\alpha$. 
Each band group can be labeled by the $C_K^d$ integers $\{\nu_{\alpha}\}$, which correspond to higher-order Chern numbers in electromagnetic responses \cite{PhysRevResearch.4.013028,PhysRevB.105.115410,petrides2018six}. 
For a particular gap, the associated qBZ is obtained as the cumulative sum of the qBZs of all band groups lying below that gap, so the gap can also associate with a $C_K^d$-element set $\{\nu_{\alpha}\}$, as a gap labeling~\cite{PhysRevResearch.4.013028,PhysRevB.105.115410,bellissard2006spaces,kellendonk2023bragg}. 

If a band group is fully occupied, the corresponding charge density is 
$n_e = S_{qBZ}/(2\pi)^d$. 
For sufficiently small $\{\phi_m\}$, the electrons evolve adiabatically within this band group, generating a current density $\bm{J}$. 
Consider an infinitesimal varation of all sublattices: $\mathbf{b}_m \to \mathbf{b}_m + \phi_m \delta \mathbf{b}$. 
For a region far from the origin, whose size is comparable to a unit cell of sublattices, the spatial phase shift $\delta\mathbf{b} \cdot\mathbf{r}$ varies negligibly across it. Within such a region, the spatial variation is equivalent to a time shift $t \to t - \delta\mathbf{b}\cdot\mathbf{r}$, 
and the two descriptions lead to the same physical response. 
Now consider a region much larger than the sublattice unit cell, the accumulated charge variation thus satisfies
\begin{eqnarray}
\delta n_eV&=&-\!\int_{\partial V} \bm{J}\!\cdot\! d\mathbf{S}\,\delta t 
= \!\int_{\mathbf{r}\in V} \nabla\!\cdot\![\bm{J}(\mathbf{r},t)\,\delta \mathbf{b}\!\cdot\!\mathbf{r}]d^d \bm{r} ,
\end{eqnarray}
where $\partial V$ is the $(d\!-\!1)$-dimensional boundary surface. $\bm{J}$ is assumed uniform in spacetime, so $\bm{J}=(2\pi)^{-d}\delta S_{qBZ}/\delta\bm{b}$ recalling $n_e = S_{qBZ}/(2\pi)^d$. Substituting the variation into Eq.~\eqref{S_BZ}, we obtain
\begin{eqnarray}
\frac{\delta S_{qBZ}}{\delta \bm{b}}&=&\sum_{{\alpha}}{\nu_{{\alpha}}} \sum_{m=1}^d \phi_{\alpha_m}\sum_{n=1}^{d}\mathcal{B}_{\alpha,mn}\bm{e}_n
\nonumber\\
&=&\frac{1}{2\pi}\sum_{{\alpha}}{\nu _{{\alpha}}}S_{{\alpha}}\sum_{m=1}^d \phi_{\alpha_m}  \bm{a}_{\alpha_m}.
\end{eqnarray}
Here $\bm{e}_{1},\bm{e}_{2},\cdots,\bm{e}_{d}$ are unit vectors in d-dimensional space. $\mathcal{B}_{\alpha,mn}$ denotes the minor of the matrix $B_\alpha$, 
whose matrix element is $(B_{\alpha})_{m,n}=\bm{b}_{\alpha_m}\cdot\bm{e}_n$.
 And $\{\bm{a}_{\alpha_m}\}$ are dual to $\{\bm{b}_{\alpha_n}\}$,
 satisfying $\bm{a}_{\alpha_m} \cdot \bm{b}_{\alpha_n} = 2\pi \delta_{m,n}$. 
 
 Let us specialize results to 1D and 2D. In 1D lattices, 
\begin{eqnarray}
    S_{qBZ}&=&\sum_{\alpha}\nu_\alpha b_\alpha,\qquad \frac{\delta S_{qBZ}}{\delta b}=\sum_{\alpha}\nu_\alpha\phi_\alpha.
    \label{1Dquasi}
\end{eqnarray}
In 2D lattices, 
\begin{eqnarray}
S_{qBZ}&=&\sum_{<\alpha_1,\alpha_2>}\nu_{\alpha_1,\alpha_2}(\bm{b_{\alpha_1}}\times\bm{b}_{\alpha_2})\cdot \bm{e}_z,\nonumber\\
\frac{\delta S_{qBZ}}{\delta \bm{b}}&=&\sum_{<\alpha_1,\alpha_2>}\nu_{\alpha_1,\alpha_2}(\phi_{\alpha_1}\bm{b}_{\alpha_2}-\phi_{\alpha_2}\bm{b}_{\alpha_1})\times\bm{e}_z.
\label{2Dquasi}
\end{eqnarray}

We now turn to the bulk-state dynamics. In the conventional single-band Thouless pump, a localized wavepacket excites a single band and evolves adiabatically. However, when the unit cell is much larger than the sublattice unit cell, a band group contains many bands separated by exponentially small gaps. Zener tunneling within the group becomes difficult to avoid \cite{wittig2005landau}, and in the quasi-periodic limit entirely unavoidable. We therefore propose a pumping scheme: the wavepacket undergoes complete intra-group tunneling while remaining adiabatic with respect to adjacent band groups. This requires the driving frequencies to satisfy~(c.f.,~App.~\ref{app_b})
\begin{eqnarray}
    \delta^2/E\ll\phi_m \ll \Delta^2/E.
    \label{eq:LZ}
\end{eqnarray}
Here $E$ is the width of the whole band group, $\delta$ is a characteristic scale of the intra-group band gaps and $\Delta$ is the gap to adjacent band groups. Under this condition, the wavepacket occupies only a single band at each moment away from the tunneling points. Note that the large unit cell tends to flatten the bands, suppressing the spreading of the wavepacket. A key question remains: does the wavepacket stay localized and drift coherently, and what determines its velocity? Now we use the 1D case to illustrate that the pumping process is governed by an effective potential emerging from couplings between adjacent band groups, and we will derive its explicit form below.

In 1D, each isolated band group $\mathrm{Band}_j$ is bounded by two Bragg planes located at $\pm G_{j-1}/2$ and $\pm G_j/2$, corresponding to two Bragg vectors $G_{j-1}=\sum_m\nu_{j-1,m}b_m$ and $G_j=\sum_m\nu_{j,m}b_m$ ($\nu_m\in\mathbb{Z}$), which define the edges of the 1D qBZ.
Within the NFE approximation, we retain only the three plane wave components 
with the strongest interband couplings, $|k\rangle$, $|k + G_{j-1}\rangle$, and $|k + G_j\rangle$, 
with $k$ lying in the interval $(-G_j/2, -G_{j-1}/2)$, as illustrated in Fig.~\ref{fig.1}(b). Then we define the three basis states as 
$\rho_1 = |k\rangle$, 
$\rho_2 = e^{-i\Phi_{j-1}t}|k + G_{j-1}\rangle$, and 
$\rho_3 = e^{-i\Phi_j t}|k + G_j\rangle$, 
where $\Phi_{j-1} = \sum_n \nu_{j-1,n}\phi_n$ and 
$\Phi_j = \sum_n \nu_{j,n}\phi_n$. Such that the reduced Hamiltonian in these basis is real,
\begin{eqnarray}
    H =
\begin{pmatrix}
\varepsilon_0 & \Delta_1 & \Delta_2 \\
\Delta_1 & \varepsilon_0 - \varepsilon_- & \Delta_3 \\
\Delta_2 & \Delta_3 & \varepsilon_0 + \varepsilon_+
\end{pmatrix},
\end{eqnarray}
where $\varepsilon_0 = k^2 / 2$ is the unperturbed energy of the state $|k\rangle$, 
while $\varepsilon_0 - \varepsilon_-$ and $\varepsilon_0 + \varepsilon_+$ correspond 
to the unperturbed energies of $|k + G_{j-1}\rangle$ and $|k + G_j\rangle$, respectively. 
See App.~\ref{app_c} for the explicit expressions of $\Delta_1$, $\Delta_2$, and $\Delta_3$.  

\begin{figure}[b]
\centering
\includegraphics[width=0.45\textwidth]{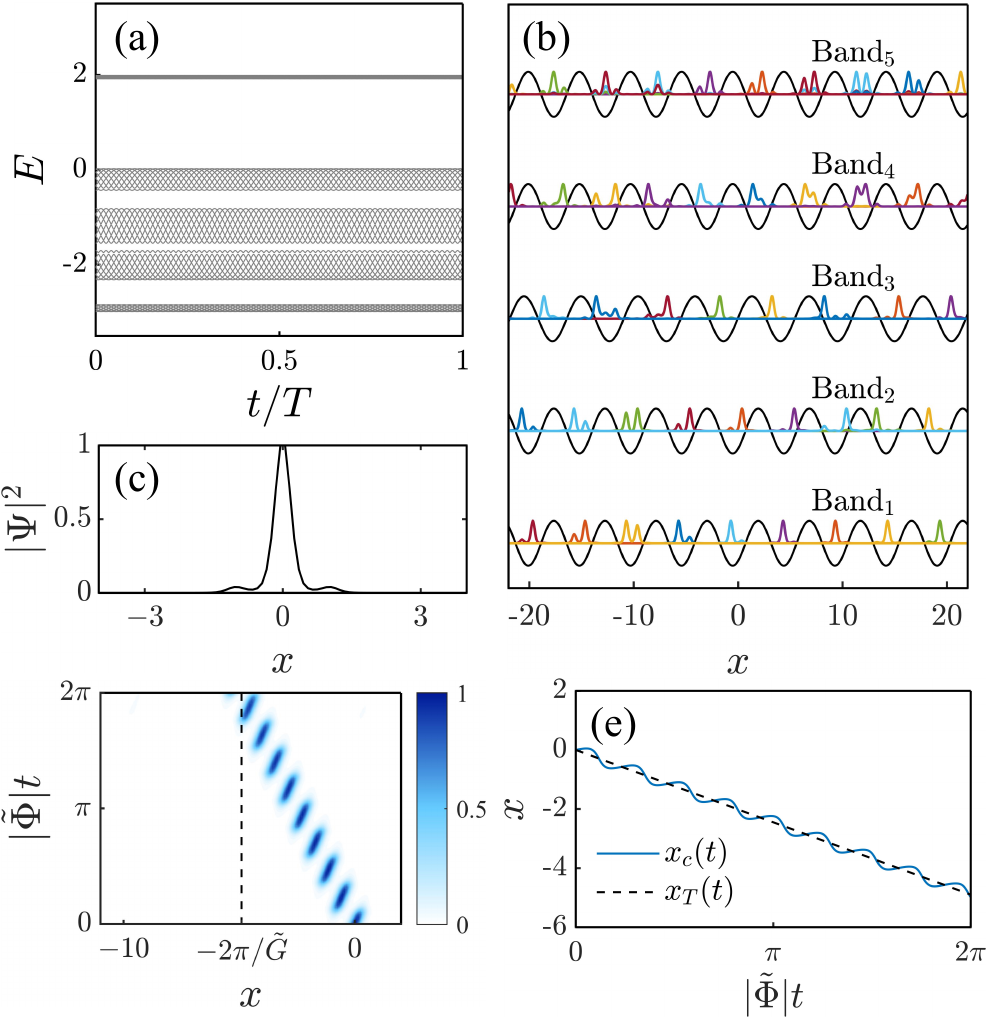}% Here is how to import EPS art
\caption{\label{fig.2} (a,b) Results of the periodic lattice. (a) Evolution of $E_{n,k=0}(t)$. (b) For each band group, $\tilde{V}(x,t)$ is plotted in black line and $|\psi_{n,k=0}(x,t)|^2$ of different bands are shown as colored lines, here we fix $\phi_1t=2$. (c-e) Results of the quasi-periodic lattice with a simulation domain of $x\in[-25,25]$ under periodic boundary conditions. (c) The instantaneous ground state at $t=0$ is the initial state. (d) Time evolutions of $|\Psi(x,t)|^2$ in the quasi-periodic lattice for excitation in $\mathrm{Band}_1$ and $\mathrm{Band}_3$, respectively. (e) COM and theoretical displacement of $|\Psi(x,t)|^2$. In all cases, $\phi_1=3\pi\times10^{-3}$, $P_1=-8$ and $P_2=-3.5$.}
\end{figure}

We denote by $\psi_j$ the eigenstate associated with the middle eigenvalue of H, which corresponds to the pertubed eigenstate of $\mathrm{Band}_j$. The momentum separation between adjacent Bragg planes, i.e.,  $\tilde{G}=G_j - G_{j-1}$, is sufficiently small, so $\Delta_3 \ll \Delta_{1,2}$ and $|\varepsilon_{\pm}| \ll \Delta_{1,2}$. 
Under this condition,
$\psi_j \simeq \mathcal{N}\,(-\kappa\rho_1-\Delta_1\rho_2+\Delta_2\rho_3)$, here $\mathcal{N}$ is the normalization factor and $\kappa \ll \Delta_{1,2}$,
with higher-order corrections in App.~\ref{app_c}.
The corresponding probability density becomes
\begin{eqnarray}
|\psi_j|^2
&=&\mathcal{N}^2[\kappa^2+\Delta_1^2+\Delta_2^2+2\kappa\Delta_1\cos(G_{j-1}x-\Phi_{j-1}t)\nonumber\\&&-2\kappa\Delta_2\cos(G_{j}x-\Phi_{j}t)-2\Delta_1\Delta_2\cos(\tilde{G}x-\tilde{\Phi}t)]\nonumber\\
&\approx&\mathcal{N}^2[\Delta_1^2+\Delta_2^2-2\Delta_1\Delta_2\cos(\tilde{G}x-\tilde{\Phi}t)].
\end{eqnarray}
Here $\tilde{\Phi}=\Phi_j-\Phi_{j-1}$. 
Thus, $\psi_j$ is well approximated by the eigenstate of an effective periodic potential
\begin{equation}
\tilde{V}(x,t)= \mathrm{sign}(\Delta_1\Delta_2)\,\mathcal{V}\cos(\tilde{G}x-\tilde{\Phi}t),
\label{V1Dequiv}
\end{equation}
where $\mathcal{V}>0$ is a scaling factor. In the strong potential regime, many plane wave components participate in the coupling. However, the harmonic term $\cos(\tilde{G}x-\tilde{\Phi}t)$ is governed by the dominant interband coupling ($\mathrm{Band}_{j-1}$ and $\mathrm{Band}_{j}$),  
so it remains the leading contribution in the eigenstates. 
Consequently, the eigenstates stays localized near the minima of $\tilde{V}(x,t)$, as shown in the next section of Fig.~\ref{fig.2}(b). 
It thus becomes clear that the drift of the initial state can be viewed from the following perspective:  
the long-period potential $\tilde{V}$ drives the state to drift with a constant velocity $\bm{v}=\tilde{\Phi}/\tilde{G}$, while the effect from other band couplings act as quasi-periodic disorder that help maintain localization. Combining it with Eq.~\eqref{1Dquasi}, we find that the drift velocity $\bm{v}$ is directly linked to the geometry of the qBZ as our main result:
\begin{equation}
    \bm{v}=\frac{1}{S_{qBZ}}\frac{\delta S_{qBZ}}{\delta \bm{b}},
    \label{velocity}
\end{equation}
which is expected to remain valid in higher dimensions, as shown in a 2D example later,
since the derivation of the $\tilde{V}(x,t)$ can be naturally generalized to arbitrary dimensions. Note that Eq.~\eqref{V1Dequiv} naturally derives the pumping current as $J_{1D}=(2\pi)^{-1}\sum_{\alpha}\nu_{\alpha}\phi_{\alpha}$, confirming that the pumping process is controlled by the effective potential.

\section{Results in commensurate
regime}
Now suppose the whole system is commensurate in spacetime, with $d$ lattice vectors $\{\bm{A}_1,\bm{A}_2,\cdots,\bm{A}_d\}$, $d$ reciprocal lattice vectors $\{\bm{B}_1,\bm{B}_2,\cdots,\bm{B}_d\}$ and temporal period $T$.
Then,
for each set $\{\bm{b}_{\alpha_1},\cdots,\bm{b}_{\alpha_d}\}$, the following relation holds:  $\bm{A}_n=\sum_{m=1}^{d}r_{\alpha,nm}\bm{a}_{\alpha_m} $, $\bm{b}_{\alpha_n}=\sum_{m=1}^{d}r_{\alpha,nm}\bm{B}_m $ and $\tau_{\alpha_m}=(\phi_{\alpha_m}T)/2\pi$, where $r_{\alpha,nm}$, $\tau_{\alpha_m}$ $\in\mathbb{Z}$. Thus the number of bands in a particular band group, denote as $N$, is given by the ratio between $S_{qBZ}$ and that of the first Brillouin zone of the underlying periodic structure:
\begin{eqnarray}
N &=&\sum_{\left< \alpha\right>}{\nu _{\alpha}}\star(\bm{b}_{\alpha_1}\land \cdots \land \bm{b}_{\alpha_d})/\star(\bm{B}_1\land\cdots \land \bm{B}_d)\nonumber\\
 &=&\sum_{\left< \alpha\right>}{\nu _{\alpha}}\det(R_{\alpha}).
 \label{No.band}
\end{eqnarray} 
The Chern number associated with the direction along $A_m$ can be expressed as the Berry curvature flux over the $(k_m,t)$ manifold \cite{vanderbilt2018berry}, 
\begin{eqnarray}
    C_m = \frac{1}{2\pi}\sum_{n\in \text{occupied bands}}\int_{\mathrm{BZ}} d^d\mathbf{k} \int_0^T dt\,f^{(n)}(\mathbf{k},t),
\end{eqnarray} 
where $f^{(n)}\equiv\partial_t a^{(n)}_{k_m}-\partial_{k_m}a^{(n)}_t$ and $a^{(n)}_\mu \equiv i\langle \psi_{n,k}(\bm{r},t)|\partial_\mu |\psi_{n,k}(\bm{r},t)\rangle$ is the Berry connection. 
It can be alternatively calculated as the total transported charge along $\bm{A}_m$ within one pumping cycle~(c.f.~App.~\ref{app_a}):
\begin{eqnarray}
C_m &=& \star(A_1 \land \cdots \land \bm{A}_{m-1} \land J_m \bm{e}_{A_m}\land \bm{A}_{m+1}\land \cdots \land A_d) T \nonumber\\
 &=&  \sum_{\left< \alpha\right>} \nu_{\alpha} \sum_{n=1}^d \tau_{\alpha_n} \mathcal{R}_{\alpha,nm}.
 \label{Chern}
\end{eqnarray} 
Here $\bm{e}_{A_m}=\bm{A}_m/|\bm{A}_m|$, and $J_m$ denotes the magnitude of the component of $\bm{J}$ projected onto $\bm{A}_m$. And $\mathcal{R}_{\alpha,nm}$ denotes the minor of the matrix $R_\alpha$ whose matrix element is $(R_{\alpha})_{n,m}=r_{\alpha,nm}$. 

In 1D lattices, 
\begin{eqnarray}
    N&=&\sum_{\alpha} \nu_\alpha r_{\alpha},\qquad C=\sum_{\alpha}\nu_\alpha\tau_\alpha.
    \label{1Dnotquasi}
\end{eqnarray}
while in 2D lattices, 
\begin{eqnarray}
N&=&\sum_{<\alpha_1,\alpha_2>}\nu_{\alpha_1,\alpha_2}(t_{\alpha,11}\cdot t_{\alpha,22}-t_{\alpha,12}\cdot t_{\alpha,21}),\nonumber\\
    C_1 &=& \sum_{<\alpha_1,\alpha_2>}\nu_{\alpha_1,\alpha_2}(r_{\alpha,22}\tau_{\alpha_1}-r_{\alpha,12}\tau_{\alpha_2}),\nonumber\\
C_2 &=& \sum_{<\alpha_1,\alpha_2>}\nu_{\alpha_1,\alpha_2}(-r_{\alpha,21}\tau_{\alpha_1}+r_{\alpha,11}\tau_{\alpha_2}).
\label{2dnotquasi}
\end{eqnarray}

In the commensurate regime, the relation between pumping current and Chern numbers in Eq.~\eqref{Chern} can be written as $\bm{J}=\sum_{m=1}^d C_m\bm{A}_m/(\Omega T)$, where $\Omega$ is the volume of the unit cell. Recall that $n_e=N/\Omega$, the expression of the drift velocity in Eq.~\eqref{velocity} then reduces to 
\begin{equation}
\bm{v}=\sum_{m=1}^{d} \frac{C_m \bm{A}_m}{NT},    
\end{equation}
as if all bands are ``averagedly'' populated observed numerically before~\cite{yang2024observation}.

\begin{figure}[b]
\centering
\includegraphics[width=0.45\textwidth]{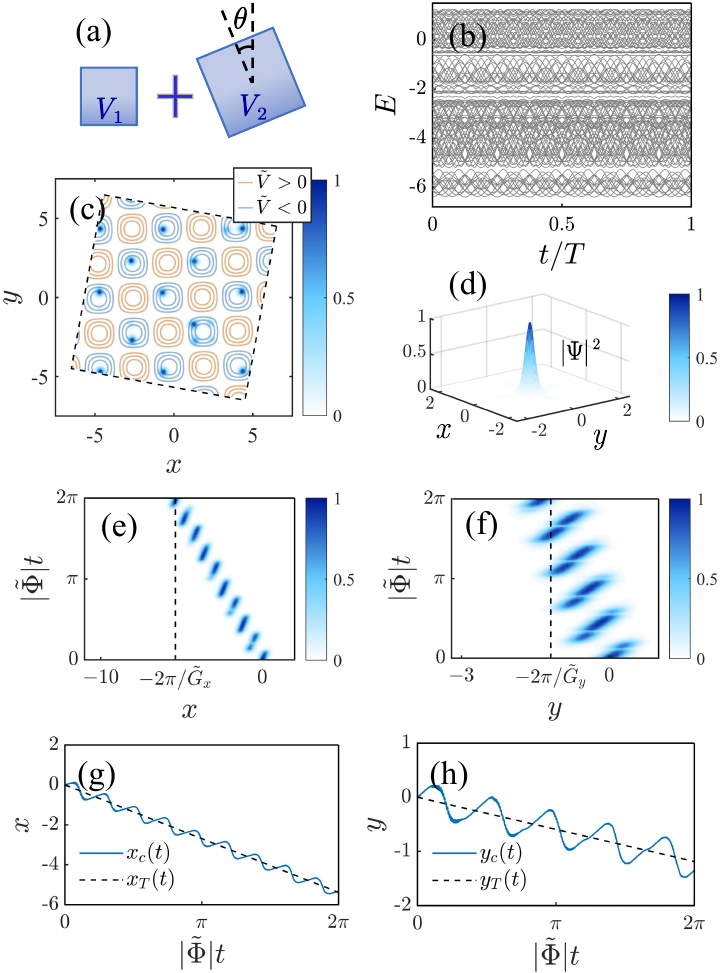}% Here is how to import EPS art
\caption{\label{fig.3}(a) Schematic of $V_{2D}(\mathbf r,t)$. (b,c) Results of the periodic lattice.
(b) Evolution of $E_{n,k=0}(t)$. (c) $|\psi_{n,k=0}(r,t)|^2$ of all bands in the $\mathrm{Band}_1$ are shown in a blue color scale. Contours of $\tilde{V}(\bm{r},t)$ in Eq.~\eqref{V2Dequiv} are overlaid. Here we fix $\phi_1t=2$. The dashed line indicates the boundary of one unit cell. (d-h) Results of the quasi-periodic lattice with a simulation domain of $[-25,25]\times [-25,25]$ under periodic boundary conditions. (d) The instantaneous ground state at $t=0$ is the initial state. (e,f) Time evolutions of $|\Psi(x,y,t)|^2$ in the quasi-periodic lattice when $\mathrm{Band}_1$ is excited; (e) and (f) show the projections along $x$ and $y$, respectively, by taking $\max(|\Psi(x,y,t)|^2)$ over the other coordinate. (g,h) COM and theoretical displacements of $|\Psi(x,y,t)|^2$ along the x and y directions. A $0.1\protect\%$ peak-density threshold is applied in 2D COM calculations to suppress numerical noise. In all cases, $\phi_1=3\pi\times10^{-3}$, $P_1=-8$ and $P_2=-4$.}
\end{figure}

\section{Pumping process in quasi-periodic lattices}

We numerically simulate the pumping dynamics in 1D and 2D lattices to confirm our general framework and effective-potential proposal.

In 1D,
we take the potential as
$V_{1D}(x,t)=P_1\cos(b_1x-\phi_1t)+P_2\cos(b_2x-\phi_2t)$. An isolated band group $\mathrm{Band}_j$ is labeled by the Bragg vector $G_j=\nu_{j,1}b_1+\nu_{j,2}b_2$, $\nu_j\in\mathbb{Z}$.  
We first consider a periodic case ($b_1=2\pi$, $b_2=\tfrac{35}{22}\pi$, $\phi_2=\tfrac{4}{3}\phi_1$), whose instantaneous band evolution is shown in Fig.~\ref{fig.2}(a).  
The lowest five gaps, from low to high energy, are labeled by $(1,-1)$, $(2,-2)$, $(-1,2)$, $(0,1)$, and $(1,0)$. The numerically obtained Chern numbers for each band group  are -1, -1, 7, -1, -1, respectively, matching Eq.~\eqref{1Dnotquasi}.  
Figure~\ref{fig.2}(b) displays effective potentials and Bloch-state densities $|\psi_{n,k=0}(x,t)|^2$ at $\phi_1 t=2$. $|\psi_{n,k=0}(x,t)|^2$ are all exactly localized near the minima of their effective potentials,
confirming their role in governing the adiabatic drift.

To demonstrate the pumping process, we introduce spacetime quasi-periodic structure by setting $b_2=5$ and $\phi_2=\tfrac{4}{\pi}\phi_1$, which shares the same gap labeling as the former periodic one. The time-dependent simulations solve Eq.~\eqref{eq:sde} by 
the time splitting spectral method \cite{BAO2002487} with periodic boundary conditions. The initial state is the instantaneous ground state at $t=0$, which excites $\mathrm{Band}_1$, as shown in Fig.~\ref{fig.2}(c). The simulation domain $x\in[-25, 25]$, much larger than the wavepacket extent, so finite-size effects are negligible. We set $\phi_1 = 3\pi\times 10^{-3}$ to satisfy Eq.~\eqref{eq:LZ}, ensuring complete intra-group tunneling and inter-group adiabaticity. From Eq.~\eqref{V1Dequiv} and the gap labeling of $\mathrm{Band}_1$, the effective potential is $\tilde{V}=-V\cos(\tilde{G}x-\tilde{\Phi}t)$, where $\tilde{G}=b_1-b_2>0$, $\tilde{\Phi}=\phi_1-\phi_2<0$ and $V>0$ is the scaling factor.  We use the sliding phase $|\tilde{\Phi}|t$ of effective potential as the time scale, and the theoretical displacement of wavepacket is $x_T(t)=-1/\tilde{G}\cdot|\tilde{\Phi}|t$. Time evolutions of $|\Psi(x,t)|^2$ in Fig.~\ref{fig.2}(d) reveal robust and localized drift. The center of mass (COM) displacement $x_c(t) = \int x|\Psi(x,t)|^2 dx/\int|\Psi(x,t)|^2 dx$ shown in Figs.~\ref{fig.2}(e) oscillates around the theoretical prediction $x_T(t)$, confirming the quantitative agreement.

We next construct a 2D potential: $V_{2D}(\mathbf r,t)=V_1+V_2$, where $V_i=P_i\cos(\bm{b}_{i1}\cdot\bm{r}-\phi_i t)+P_i\cos(\bm{b}_{i2}\cdot\bm{r}-\phi_i t)$, with $i=1,2$. Four reciprocal lattice vectors are $\bm{b}_{11}=(\beta_1,0)$, $\bm{b}_{12}=(0,\beta_1)$ and $\bm{b}_{21}=(\beta_2\cos\theta,\beta_2\sin\theta)$, $\bm{b}_{22}=(-\beta_2\sin\theta,\beta_2\cos\theta)$ illustrated in Fig.~\ref{fig.3}(a). We first take a periodic structure with parameters  $\theta=\arctan(7/24)$, $\beta_1=2\pi$, $\beta_2=\frac{8}{5}\pi$ and $\phi_2=4/3$. The instantaneous band evolution is shown in Fig.~\ref{fig.3}(b). For the lowest isolated band group ($\mathrm{Band}_1$), the Bragg vectors are $\tilde{\bm{G}}_1=\bm{b}_{11}-\bm{b}_{21}$ and $\tilde{\bm{G}}_2=\bm{b}_{12}-\bm{b}_{22}$. Two Chern numbers are $C_1=-5$ and $C_2=-1$,
equal to those calculated from Eq.~\eqref{2dnotquasi}. The associated effective potential is  
\begin{eqnarray}
    \tilde{V}(\bm{r},t)=-\mathcal{V}\cos(\tilde{\bm{G}}_1\cdot\bm{r}-\tilde{\Phi}t)-\mathcal{V}\cos(\tilde{\bm{G}}_2\cdot\bm{r}-\tilde{\Phi}t),
    \label{V2Dequiv}
\end{eqnarray}
where $\tilde{\Phi}=\phi_1-\phi_2<0$, and $\mathcal{V}>0$ is the scaling factor. As shown in Fig.~\ref{fig.3}(c), the Bloch-state densities of $\mathrm{Band}_1$ at $\phi_1 t=2$ are strongly localized near the minima of $\tilde V$. Then we  set $\beta_2=5$, $\theta=10^\circ$, and $\phi_2=\tfrac{4}{\pi}\phi_1$ to make $V_{2D}(\bm{r},t)$ be spacetime quasi-periodic. From Eq.~(\ref{V2Dequiv}), the theoretical displacement of wavepacket is $\bm{r}_T(t)=(x_T(t),y_T(t))=-(\tilde{\bm{G}_1}+\tilde{\bm{G}_2})/|\tilde{\bm{G}_1}|^2\cdot|\tilde{\Phi}|t$, the same result is available from Eq.~\eqref{velocity}. Let $\xi_T(t)=-\tilde{G}^{-1}_\xi\cdot|\tilde{\Phi}|t\,(\xi=x,y) $, then
\begin{eqnarray}
  {\tilde{G}_x}^{-1}=\frac{\beta_1+\beta_2(\sin\theta-\cos\theta)}{\beta_1^2+\beta_2^2-2\beta_1\beta_2\cos\theta},\\
{\tilde{G}_y}^{-1}=\frac{\beta_1-\beta_2(\sin\theta+\cos\theta)}{\beta_1^2+\beta_2^2-2\beta_1\beta_2\cos\theta}.
\end{eqnarray}
 We set $\phi_1=3\pi\times10^{-3}$ and use the ground state at $t=0$ to excite $\mathrm{Band}_1$, as shown in Fig.~\ref{fig.3}(d). We take the simulation domain as $[-25, 25]\times[-25, 25]$.
 Time evolution of $|\Psi(\bm{r},t)|^2$ are shown in Figs.~\ref{fig.3}(e,f), and the COM displacements $\xi_c(t) = \int \xi|\Psi|^2 d^2\bm{r}/\int|\Psi|^2 d^2\bm{r}$ ($\xi=x,y$) are shown in Figs.~\ref{fig.3}~(g,h), in an excellent consistency with our proposal.

\section{Conclusion}
We have developed a universal framework for Thouless pumps in continuous models with spacetime quasi-periodicity. We show that the evolution of a localized state is governed by a long-wavelength effective potential emerging from interband couplings. As a consequence, we establish a universal relation between the topological drift velocity and the geometric structure of the qBZ. Numerical simulations in 1D and 2D quasi-periodic lattices confirm the theory and its dynamical picture, revealing stable, localized and directional drift of an initially localized state. Our approach of analyzing the dynamics from the perspective of an effective potential may offer inspiration for studying interacting Thouless pumps involving multiple bands as future interest.

\begin{acknowledgements}
The authors thank Fangwei Ye for useful discussions.
The work of Y. Y. was supported by the National Key Research and Development Program of China (Grant No. 2024YFA1408303), the National Natural Science Foundation of China (Grants No. 12474157 and No. 12447103), the sponsorship from Yangyang Development Fund, and Xiaomi Young Scholars Program.
\end{acknowledgements}

\section*{Data Availability} The data that support the findings of this study are available from the authors upon reasonable request.

\appendix

\section{Detailed calculations of several quantities}\label{app_a}
In this section, we present the full derivation of the variation $\frac{\delta S_{qBZ}}{\delta \bm{b}}$ and the Chern numbers $C_m$.
We first perform the variation $\mathbf{b}_m \to \mathbf{b}_m + \phi_m \delta \mathbf{b}$ to calculate the response of $S_{qBZ}$. Recalling that $S_{qBZ} = \sum_{{\alpha}}{\nu _{{\alpha}}}S_{\alpha}$, where $S_{\alpha}=\star(\bm{b}_{\alpha_1}\land \bm{b}_{\alpha_2}\land \cdots \land \bm{b}_{\alpha_d})$, 
then we obtain
\begin{eqnarray}
&&\delta S_{qBZ} \nonumber\\
&=&\!\!\!\sum_{ \alpha}{\nu _{\alpha}}\star[( \bm{b}_{\alpha_1}+\phi _{\alpha_1}\delta \bm{b} ) \land \cdots \land (\bm{b}_{\alpha_d}+\phi _{\alpha_d}\delta \bm{b})]-S_{qBZ} \nonumber\\
&=& \sum_{ \alpha}{\nu _{\alpha}}\sum_{m=1}^d\left( -1 \right) ^{m-1}\phi _{\alpha_m}\delta \bm{b}\cdot\nonumber\\
&&\star( \bm{b}_{\alpha_1}\land \cdots \land \bm{b}_{\alpha_{m-1}}\land \bm{b}_{\alpha_{m+1}}\land \cdots \land \bm{b}_{\alpha_d}).
\label{J1}
\end{eqnarray}
Consequently,
\begin{eqnarray}
&&\frac{\delta S_{qBZ}}{\delta \bm{b}}\nonumber\\
 &=& \sum_{ \alpha}{\nu _{\alpha}} \sum_{m=1}^d \phi_{\alpha_m} \det \left( 
\begin{matrix}
b_{\alpha_1}^{1} & b_{\alpha_1}^{2} & \cdots & b_{\alpha_1}^{d} \\
\vdots & & & \vdots \\
b_{\alpha_{m-1}}^{1} & b_{\alpha_{m-1}}^{2} & \cdots & b_{\alpha_{m-1}}^{d} \\
\bm{e}_{1} & \bm{e}_{2} & \cdots & \bm{e}_{d} \\
b_{\alpha_{m+1}}^{1} & b_{\alpha_{m+1}}^{2} & \cdots & b_{\alpha_{m+1}}^{d} \\
\vdots & & & \vdots \\
b_{\alpha_d}^{1} & b_{\alpha_d}^{2} & \cdots & b_{\alpha_d}^{d},
\end{matrix}
\right)\nonumber\\
&=& \sum_{{\alpha}}{\nu_{{\alpha}}} \sum_{m=1}^d \phi_{\alpha_m}\sum_{n=1}^{d}\mathcal{B}_{\alpha,mn}\bm{e}_n
\end{eqnarray}

$\bm{e}_{1},\bm{e}_{2},\cdots,\bm{e}_{d}$ are unit vectors in d-dimensional space, $\mathcal{B}_{\alpha,mn}$ denotes the minor of the matrix $B_\alpha$, whose matrix element is $(B_{\alpha})_{m,n}=\bm{b}_{\alpha_m}\cdot\bm{e}_n$.We now introduce a set of dual lattice vectors $\{\bm{a}_{\alpha_m}\}$ via
\begin{eqnarray}
&&\left( -1 \right) ^{m-1}\star( \bm{b}_{\alpha_1}\land \cdots \land \bm{b}_{\alpha_{m-1}}\land \bm{b}_{\alpha_{m+1}}\land \cdots \land \bm{b}_{\alpha_d})\nonumber\\ &=& \frac{1}{2\pi}S_{\alpha_1,\alpha_2,\cdots ,\alpha_d}\bm{a}_{\alpha_m}
\end{eqnarray}
These dual vectors satisfy $\bm{a}_{\alpha_m} \cdot \bm{b}_{\alpha_n} = 2\pi \delta_{m,n}$, then we obtain
\begin{eqnarray}
\frac{\delta S_{qBZ}}{\delta \bm{b}} = \frac{1}{2\pi}\sum_{{\alpha}}{\nu _{{\alpha}}}S_{\alpha}\sum_{m=1}^d \phi_{\alpha_m}  \bm{a}_{\alpha_m}.
\end{eqnarray}

Now we work in the commensurate regime and use the same notation $\bm{A}_n=\sum_{m=1}^{d}r_{\alpha,nm}\bm{a}_{\alpha_m} $, $\bm{b}_{\alpha_n}=\sum_{m=1}^{d}r_{\alpha,nm}\bm{B}_m $ and $\tau_{\alpha_m}=(\phi_{\alpha_m}T)/2\pi$ as in the main text. $\{\bm{A}_n\}$ and $\{\bm{B}_m\}$ denote the lattice vectors and their reciprocal counterparts, respectively. T is the pumping period.
 The Chern number $C_m$ of a Thouless pump is defined as the total amount of charge transported along a given direction $\bm{A}_m$ during one pumping cycle,
\begin{eqnarray}
&&C_m\nonumber\\
 &=& \star(A_1 \land \cdots \land \bm{A}_{m-1} \land J_m \bm{e}_{A_m}\land \bm{A}_{m+1}\land \cdots \land A_d) T \nonumber\\
&=& (-1)^{m-1}\frac{1}{2\pi} (\bm{J}\cdot\bm{B}_m)\bm{A}_m \cdot \star( \bm{A}_1 \land \bm{A}_2 \land \cdots \land \bm{A}_d) T \nonumber\\
&=& \frac{\Omega}{2\pi}\bm{J}\cdot\bm{B}_m T.
\label{C1}
\end{eqnarray}
Here $\bm{e}_{A_m}=\bm{A}_m/|\bm{A}_m|$ and $J_m$ denotes the magnitude of the component of $\bm{J}$ projected onto $\bm{e}_{\bm{A}_m}$. $\Omega$ is the unit cell volume. Using the relation $\bm{J}=(2\pi)^{-d}\delta S_{qBZ}/\delta \bm{b}$ and substituting Eq.~\eqref{J1} into Eq.~\eqref{C1}, we obtain
\begin{widetext}
\begin{eqnarray}
C_m&=& \frac{\Omega}{(2\pi)^d} \sum_{ \alpha}{\nu _{\alpha}} \sum_{n=1}^d \tau_{\alpha_n} \star( \bm{b}_{\alpha_1} \land \cdots \land \bm{b}_{\alpha_{n-1}} \land \bm{B}_m \land \bm{b}_{\alpha_{n+1}} \land \cdots \land \bm{b}_{\alpha_d}) \nonumber\\
&=& \sum_{ \alpha} \nu_{\alpha} \sum_{n=1}^d \tau_{\alpha_n}
\det\left(
\begin{matrix}
r_{\alpha,1,1} & \cdots & r_{\alpha,1,m-1} & r_{\alpha,1,m} & r_{\alpha,1,m+1} & \cdots & r_{\alpha,1,d} \\
\vdots & & & & & & \vdots \\
r_{\alpha,n-1,1} & \cdots & r_{\alpha,n-1,m-1} & r_{\alpha,n-1,m} & r_{\alpha,n-1,m+1} & \cdots & r_{\alpha,n-1,d} \\
0 & \cdots & 0 & 1 & 0 & \cdots & 0 \\
r_{\alpha,n+1,1} & \cdots & r_{\alpha,n+1,m-1} & r_{\alpha,n+1,m} & r_{\alpha,n+1,m+1} & \cdots & r_{\alpha,n+1,d} \\
\vdots & & & & & & \vdots \\
r_{\alpha,d,1} & \cdots & r_{\alpha,d,m-1} & r_{\alpha,d,m} & r_{\alpha,d,m+1} & \cdots & r_{\alpha,d,d} \\

\end{matrix}
\right) \nonumber\\
&=&  \sum_{\left< \alpha\right>} \nu_{\alpha} \sum_{n=1}^d \tau_{\alpha_n} \mathcal{R}_{\alpha,nm}.
\end{eqnarray} 
\end{widetext}
$\mathcal{R}_{\alpha,nm}$ denotes the minor of the matrix $R_\alpha$, whose matrix element is $(R_{\alpha})_{n,m}=r_{\alpha,nm}$. 

\section{Restriction on driving frequencies}\label{app_b}
\begin{figure}[b] 
\centering
\includegraphics[width=0.43\textwidth]{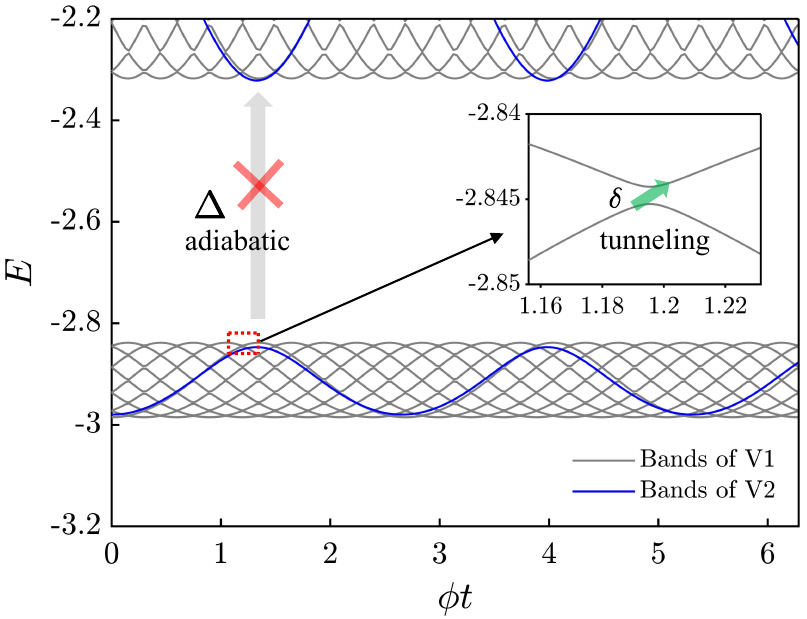}
\caption{\label{fig_A} 
The energy bands $E_{n,k=0}(t) $ of $V_1$ and $V_2$ are plotted in grey and blue lines, respectively. The inset shows a zoom of the bands from $V_1$. $\delta$ is the gap between minibands in the first band group, and the green arrow indicates complete tunneling supported in the pump scheme. $\Delta$ is the gap with respect to adjacent band group, with the gray prohibited arrow represents no tunneling to adjacent band group.}
\end{figure}

We derive the restriction $\delta^2/E\leq\phi_m\leq\Delta^2/E$ required for complete intra-group tunneling and no inter-group tunneling. The Zener tunneling probability at an avoided crossing is $P=\exp(-\pi \epsilon^2/2\beta)$ \cite{wittig2005landau}, here $\epsilon$ is the gap at the tunneling point and $\beta=|dE/dt|$ is the rate of energy variation. To estimate $\beta$, we consider two potentials: 
\begin{eqnarray}
 V_1&=&-8\cos(2\pi x-\phi t)-3.5\cos\left(\frac{35}{22}\pi x- \frac{4}{\pi}\phi t\right),\nonumber\\
  V_2&=&-8\cos(2\pi x-\phi t)-3.5\cos\left(\frac{8}{5}\pi x- \frac{4}{\pi}\phi t\right),\nonumber
\end{eqnarray}
whose bands are shown in Fig.~\ref{fig_A}. The $V_1$ possesses a multiple-band structure, and is the focus of our pumping scheme, while $V_2$  serves as an auxiliary reference. We use subscripts 1 and 2 to represent the physical quantities corresponding to $V_1$ and $V_2$. Since $V_1$ and $V_2$ differ only slightly ($35/22-8/5\approx 0.009$), the energy bands of $V_1$ can be regarded as folded energy bands of $V_2$, and $|dE/dt|_1\approx|dE/dt|_2$. From the band structure of $V_2$, one observes that $|dE/dt|_2$ varies slowly in time and can be estimated as $E\phi$. Here $E$ is the width of the first band group. Moreover, the energy variation rates for intra-group and inter-group tunneling in $V_1$ are of the same order, since both originate from the same underlying band dispersion of $V_2$. The intra-group and inter-group tunneling probabilities are then approximately $\exp[-\delta^2/(E\phi)]$ and $\exp[-\Delta^2/(E\phi)]$, respectively. Complete intra-group tunneling requires $\delta^2/E\phi\ll1$, while suppression of inter-group tunneling requires $\Delta^2/E\phi\gg 1$,  yielding $\delta^2/E\ll\phi\ll\Delta^2/E$. Extending to an arbitrary multi-frequency system, we obtain $\delta^2/E\ll\phi_m\ll\Delta^2/E$.

\section{Effective Hamiltonian and effective potential}\label{app_c}
In this section, we construct the effective Hamiltonian matrix and derive the corresponding effective potential under the NFE approximation in 1D. We consider the j-th isolated group of bands (denote as $\operatorname{Band}_j$), in which one of the unperturbed eigenstates is $\psi_j^0=|k\rangle=\frac{1}{\sqrt{L}}\exp(ikx)$. The Dyson--Schwinger equation \cite{sakurai2020modern} leads to an effective Hamiltonian acting on the low-energy subspace,
\begin{eqnarray}
&&H_{\mathrm{eff}}(E) \nonumber\\
&=& PHP + PHQ(E - QHQ)^{-1}QHP,\nonumber\\
&=&PH_0P + \sum_{n=0}^{\infty}PV(Q\frac{1}{E - H_0}QV)^n P. 
\label{expand}
\end{eqnarray}
where $P$ projects onto the selected subspace that we care about, and $Q = 1 - P$ projects onto the remaining states.  The perturbed eigenfunction is $H_{\mathrm{eff}}(E)\psi_j=E\psi_j$, where we have restricted $\psi_j$ in the selected subspace. The expansion occurs when there exist energy separation between $P$ and $Q$. In our model, the time-dependent potential takes the form $V(x,t) = \sum_m V_m e^{i(b_m x - 2\pi \phi_m t)} + \text{c.c.}$. Thus, the two unperturbed states $|k_a\rangle=\frac{1}{\sqrt{L}}\exp(ik_\alpha x)$ and $|k_{\beta}
\rangle=\frac{1}{\sqrt{L}}\exp(ik_\beta x)$ can be coupled only when $k_\beta-k_\alpha=\sum\nu_m b_m$, $\nu_m\in \mathbb{Z}$. Their coupling coefficient is
\begin{eqnarray}
&&\langle k_\alpha|H_{\mathrm{eff}}(E)|k_\beta\rangle\nonumber\\
&=&
\sum_{p=(p_1,\dots,p_l)}
\frac{\prod_{j=1}^{l}V_{p_j}\exp(-i\phi_{p_j} t)}{\prod_{j=1}^{l-1}E-\frac{1}{2}(k_\alpha+\sum_{s=1}^{j}b_{p_s})^2}.
\end{eqnarray}
Here $b_{p_{1,2,\cdots,l}}\in\{b_m\}$, and the summation runs over all possible scattering paths $p=\{p_1,\dots,p_l\}$ that connect the two momenta 
$k_\alpha$ and $k_\beta$, satisfying $\sum_{s=1}^{l} b_{p_s}=k_\beta-k_\alpha$. So the accumulated phase is the same for each path $p$ ,i.e., $\sum_{j=1}^{l} \phi_{p_j} t
= \sum_m \nu_m \phi_m t$. We can thus denote the coupling coefficients as $\langle k_\alpha|H_{\mathrm{eff}}(E)|k_\beta\rangle=e^{-i\sum_m \nu_m \phi_m t}\Delta_{\alpha\beta}$, where $\Delta_{\alpha\beta}\in\mathbb{R}$.

In 1D, $\operatorname{Band}_j$ is bounded by two Bragg planes 
located at $\pm G_{j-1}/2$ and $\pm G_j/2$, corresponding respectively to 
the lower $\operatorname{Gap}_{j-1}$ and the upper $\operatorname{Gap}_j$, where $G_j=\sum\nu_{j,m}b_m$ and $G_{j-1}=\sum\nu_{j-1,m}b_m$. We set $k\in(-G_j/2,-G_{j-1}/2)$ and retain only the two states that couple most strongly to $|k\rangle$, namely $|k + G_{j-1}\rangle$ in $\operatorname{Band}_{j-1}$ and $|k + G_j\rangle$ in $\operatorname{Band}_{j+1}$. So
\begin{eqnarray}
    P=|k\rangle\langle k|+|k+G_{j-1}\rangle\langle k+G_{j-1}|+|k+G_{j}\rangle\langle k+G_{j}|.\nonumber
\end{eqnarray} 
In the self-consistent equation Eq.~\eqref{expand}, the difference between the exact eigenenergy $E$ and the unperturbed energy $\epsilon_0=\frac{1}{2}k^2$ contributes only to higher-order corrections. So we set $E=\epsilon_0$ as an approximate value when constructing the expression of $H_{\mathrm{eff}}$. 
We redefine the three basis states as
$|\rho_1\rangle = |k\rangle$, 
$|\rho_2\rangle = e^{-i\Phi_{i-1}t}|k + G_{j-1}\rangle$, and 
$|\rho_3\rangle = e^{-i\Phi_i t}|k + G_j\rangle$, 
where the phase factors are given by 
$\Phi_{j-1} = \sum_m \nu_{j-1,m}\phi_m$ and 
$\Phi_j = \sum_m \nu_{j,m}\phi_m$. 
Under such choice, all matrix elements of $H$ are real. The effective Hamiltonian matrix takes the form
\begin{align}
&H_{\mathrm{eff}}
=\varepsilon_0+
\begin{pmatrix}
0 & \Delta_1 & \Delta_2 \\
\Delta_1 & \varepsilon_- & \Delta_3 \\
\Delta_2 & \Delta_3 & \varepsilon_+
\end{pmatrix},
\end{align}
where
\begin{align}
&\left\{
\begin{array}{lcl}
\Delta_1 &=& e^{i\Phi_{j-1}t}\langle k|H_{\mathrm{eff}}|k+G_{j-1}\rangle,\\[4pt]
\Delta_2 &=& e^{i\Phi_{j}t}\langle k|H_{\mathrm{eff}}|k+G_{j}\rangle,\\[4pt]
\Delta_3 &=& e^{i(\Phi_j-\Phi_{j-1})t}\langle k+G_{j-1}|H_{\mathrm{eff}}|k+G_{j}\rangle.
\end{array}
\right.
\label{Heff_matrix}
\end{align}
Here  $\varepsilon_0 +\varepsilon_-=\frac{1}{2}(k+G_{j-1})^2$ and $\varepsilon_0 + \varepsilon_+=\frac{1}{2}(k+G_j)^2$.

Let $\cos(\theta)=\Delta_1/R$ and $\sin(\theta)=\Delta_2/R$, here $R=\sqrt{\Delta_1^2+\Delta_2^2}$. We perform a linear transformation: $|\rho_2'\rangle=\cos(\theta)|\rho_2\rangle+\sin(\theta)|\rho_3\rangle$, $|\rho_3'\rangle=-\sin(\theta)|\rho_2\rangle+\cos(\theta)|\rho_3\rangle$. The transformed Hamiltonian matrix takes the following form:
\begin{eqnarray}
H'_{\mathrm{eff}} =
\varepsilon_0+\left(
\begin{array}{ccc}
0 & R & 0 \\[4pt]
R & \varepsilon'_- & \Delta'_3 \\[4pt]
0 & \Delta'_3 & \varepsilon'_+
\end{array}
\right), 
\end{eqnarray}
where
\begin{eqnarray}
\left\{
\begin{aligned}
\varepsilon'_- &= \frac{\Delta_1^2 \varepsilon_- + \Delta_2^2 \varepsilon_+ + 2\Delta_1 \Delta_2 \Delta_3}{R^2},\\[6pt]
\varepsilon'_+ &= \frac{\Delta_2^2 \varepsilon_- + \Delta_1^2 \varepsilon_+ - 2\Delta_1 \Delta_2 \Delta_3}{R^2},\\[6pt]
\Delta'_3 &= \frac{(\Delta_1^2 - \Delta_2^2)\Delta_3 + \Delta_1 \Delta_2 (\varepsilon_+ - \varepsilon_-)}{R^2}.
\end{aligned}
\right.
\end{eqnarray}
Since $G_j - G_{j-1}$ represents the separation between two adjacent gaps in $k$-space and is relatively small, 
a sufficiently strong lattice potential may make $|\varepsilon_{\pm}|$ much smaller than $\Delta_{1,2}$. 
Moreover, the coupling $\Delta_3$ usually originates from longer scattering paths than $\Delta_{1,2}$, 
so $\Delta_3$ is also much smaller than $\Delta_{1,2}$. 
As a result, $|\varepsilon'_{\pm}|$ and $\Delta'_3$ are much smaller than $R$. Therefore, \( H'_{\mathrm{eff}} \) has two eigenvalues with magnitudes around $\varepsilon_0\pm R$, 
and one intermediate eigenvalue close to $\varepsilon_0+\varepsilon'_+$, which corresponds to the eigenenergy of \( \psi_j \). We set the eigenvalue as $\lambda = \varepsilon_0+\varepsilon'_+ + \xi$. 
Neglecting higher-order terms of $\xi$ in the eigenvalue equation yields
\begin{eqnarray}
    \xi\approx -\frac{(\Delta'_3)^2}{D}, 
\qquad 
D = \frac{R^2}{\varepsilon'_+} + \varepsilon'_- - \varepsilon'_+ .
\end{eqnarray}
Based on this eigenvalue \( \lambda \), we obtain the corresponding eigenstate 
and express it in the original basis $\{|\rho_1\rangle, |\rho_2\rangle, |\rho_3\rangle\}$ as
\begin{eqnarray}
    |\psi_j\rangle &\approx& 
\mathcal{N}(a|\rho_1\rangle+b|\rho_2\rangle-c|\rho_3\rangle),\nonumber\\
a&=&\frac{\Delta'_3}{1-\varepsilon'_{+}(\varepsilon'_{+}-\varepsilon'_{-})/R^2},\nonumber\\
b&=&\Delta_2+\frac{\Delta'_3\Delta_1}{D},\nonumber\\
c&=&\Delta_1-\frac{\Delta'_3\Delta_2}{D}.
\end{eqnarray}
where \( \mathcal{N} \) is the normalization factor. So 
\begin{eqnarray}
    |\psi_j|^2&=&\mathcal{N}^2[a^2+b^2+c^2+2ab\cdot\cos(G_{j-1}x-\Phi_{j-1}t)\nonumber\\
    &-&2ac\cdot\cos(G_{j}x-\Phi_{j}t)-2bc\cdot\cos(\tilde{G}x-\tilde{\Phi}t)].\nonumber
\end{eqnarray}
Here $\tilde{G}=G_{j}-G_{j-1}$, $\tilde{\Phi}=\Phi_j-\Phi_{j-1}$. $a$ is much smaller than $b,c$, so $\psi_j$ as the eigenstate of $\mathrm{Band}_j$ is well approximated
by the first-band eigenstate of the effective potential 
\begin{equation}
\tilde{V}(x,t)= \mathrm{sign}[(\Delta_2+\frac{\Delta'_3\Delta_1}{D})(\Delta_1-\frac{\Delta'_3\Delta_2}{D})]\cdot \mathcal{V}\cos(\tilde{G}x-\tilde{\Phi}t),
\label{V1Dequiv}
\end{equation}
$\mathcal{V}>0$ is the scaling factor. Under a wide range of parameters, $\mathrm{sign}(bc)=\mathrm{sign}(\Delta_1\Delta_2)$. If $\mathrm{Band}_j$ is bounded by extremely higher order perturbations, the difference among $\varepsilon_{\pm},\Delta_3$ and $\Delta_{1,2}$ will decrease. In very extreme cases, it may lead to $\mathrm{sign}(bc)=-\mathrm{sign}(\Delta_1\Delta_2)$.

\nocite{sakurai2020modern} 
%\bibliography{bib}


\end{document}